
\magnification\magstep1

\openup 1\jot

\input mssymb
\def\hbar{\mathchar '26\mkern -9muh}

\catcode`@=11
\def\eqaltxt#1{\displ@y \tabskip 0pt
  \halign to\displaywidth {%
    \rlap{$##$}\tabskip\centering
    &\hfil$\@lign\displaystyle{##}$\tabskip\z@skip
    &$\@lign\displaystyle{{}##}$\hfil\tabskip\centering
    &\llap{$\@lign##$}\tabskip\z@skip\crcr
    #1\crcr}}
\def\eqallft#1{\displ@y \tabskip 0pt
  \halign to\displaywidth {%
    $\@lign\displaystyle {##}$\tabskip\z@skip
    &$\@lign\displaystyle{{}##}$\hfil\crcr
    #1\crcr}}
\catcode`@=12 

\def\half{{\textstyle {1 \over 2}}}

\def\pmb#1{\setbox0=\hbox{#1}  \kern-.025em\copy0\kern-\wd0
  \kern0.05em\copy0\kern-\wd0  \kern-.025em\raise.0433em\box0 }
\def\pmbh#1{\setbox0=\hbox{#1} \kern-.12em\copy0\kern-\wd0
	    \kern.12em\copy0\kern-\wd0\box0}
\def\sqr#1#2{{\vcenter{\vbox{\hrule height.#2pt
      \hbox{\vrule width.#2pt height#1pt \kern#1pt
	 \vrule width.#2pt}
      \hrule height.#2pt}}}}

\def\rchi{{\raise 2pt \hbox {$\chi$}}}
\def\rga{{\raise 2pt \hbox {$\gamma$}}}
\def\rg{{\raise 2 pt \hbox {$g$}}}

\def\({\left(}
\def\){\right)}
\def\<{\left\langle}
\def\>{\right\rangle}

\def\[{\left[}
\def\]{\right]}
\let\text=\hbox

\def\rta{\rightarrow}

\def\ol{\overline}

\def\A{\hbox{$ A\kern -5.5pt / \kern +5.5pt$}}
\def\B{\hbox{$ B\kern -5.5pt / \kern +5.5pt$}}
\def\C{\hbox{$ C\kern -5.5pt / \kern +5.5pt$}}
\def\D{\hbox{$ D\kern -5.5pt / \kern +5.5pt$}}
\def\E{\hbox{$ E\kern -5.5pt / \kern +5.5pt$}}
\def\F{\hbox{$ F\kern -5.5pt / \kern +5.5pt$}}
\def\G{\hbox{$ G\kern -5.5pt / \kern +5.5pt$}}
\def\H{\hbox{$ H\kern -5.5pt / \kern +5.5pt$}}
\def\I{\hbox{$ I\kern -5.5pt / \kern +5.5pt$}}
\def\Z{\hbox{$ Z\kern -5.5pt / \kern +.5pt$}}

\hfuzz 6pt

\catcode`@=12 

\font\twelverm=cmr12
\font\twelvebf=cmbx12
\def\bigtype{\twelverm \twelvebf \baselineskip=16pt}

\vskip .2 true in
\centerline {\bigtype The Case of the  }
\centerline {\bigtype Missing Wormhole State\footnote*{\sevenrm Based on a
essay
which received a Honorable Mention in the  1995
Gravity Research Foundation Awards.}\footnote{,**}{\sevenrm  Talk given at the
6th
Moskow Quantum Gravity Seminar, Moskow 12-19 June 1995, Russia }}
\vskip .1 true in
\centerline {{\rm P.V. Moniz}\footnote{***}{{\sevenrm e-mail address:
prlvm10@amtp.cam.ac.uk}}}
\vskip .1 true in
\centerline {Department of Applied Mathematics and Theoretical Physics}
\centerline{
University of Cambridge}
\centerline{
 Silver Street, Cambridge,
CB3 9EW, UK }

\vskip .2 true in
\centerline {\bf ABSTRACT}
\vskip .1 true in

{\sevenrm
The issue concerning the existence
of wormhole states in locally supersymmetric minisuperspace
models with matter is addressed.
Wormhole  states
are apparently  absent in models obtained from
the more  general theory of N=1 supergravity with supermatter.
A Hartle-Hawking type solution can be found, even though some terms (which are
scalar field
dependent)
cannot be determined in a satisfactory way.
A possible cause  is   investigated here. As far as the wormhole situation is
concerned, we argue here
that the type
of Lagrange multipliers and fermionic derivative ordering one uses may make a
difference.
A proposal is made for  supersymmetric quantum wormholes to also
be invested with a Hilbert space structure, associated with
a maximal analytical extension of the corresponding
minisuperspace.}
\noindent

\magnification\magstep1

\openup 1\jot

\input mssymb
\def\hbar{\mathchar '26\mkern -9muh}

\catcode`@=11
\def\eqaltxt#1{\displ@y \tabskip 0pt
  \halign to\displaywidth {%
    \rlap{$##$}\tabskip\centering
    &\hfil$\@lign\displaystyle{##}$\tabskip\z@skip
    &$\@lign\displaystyle{{}##}$\hfil\tabskip\centering
    &\llap{$\@lign##$}\tabskip\z@skip\crcr
    #1\crcr}}
\def\eqallft#1{\displ@y \tabskip 0pt
  \halign to\displaywidth {%
    $\@lign\displaystyle {##}$\tabskip\z@skip
    &$\@lign\displaystyle{{}##}$\hfil\crcr
    #1\crcr}}
\catcode`@=12 

\def\half{{\textstyle {1 \over 2}}}

\def\pmb#1{\setbox0=\hbox{#1}  \kern-.025em\copy0\kern-\wd0
  \kern0.05em\copy0\kern-\wd0  \kern-.025em\raise.0433em\box0 }

\def\pmbh#1{\setbox0=\hbox{#1} \kern-.12em\copy0\kern-\wd0
	    \kern.12em\copy0\kern-\wd0\box0}
\def\sqr#1#2{{\vcenter{\vbox{\hrule height.#2pt
      \hbox{\vrule width.#2pt height#1pt \kern#1pt
	 \vrule width.#2pt}
      \hrule height.#2pt}}}}

\def\rchi{{\raise 2pt \hbox {$\chi$}}}
\def\rga{{\raise 2pt \hbox {$\gamma$}}}
\def\rrho{{\raise 2pt \hbox {$\rho$}}}

\def\({\left(}
\def\){\right)}
\def\<{\left\langle}
\def\>{\right\rangle}

\def\[{\left[}
\def\]{\right]}
\let\text=\hbox

\def\rta{\rightarrow}

\def\ol{\overline}


Mistery stories seem to be a must in Britain. One just has to remember famous
characters such
as Sherlock Holmes, Hercule Poirot and Miss Marple and celebrated authors like
Sir. A.C. Doyle and Agatha Christie. Furthermore, there is even a book entitled
``Cambridge Colleges Ghosts'' [0]. Hence, I hope   that the
title of this talk does
 not seem so strange after all. Let me then begin by some introdutory remarks
concerning our {\it mistery} case.

 A quantum theory of gravity
constitutes one of the foremost aspirations  in
theoretical physics
[1].
The inclusion of    supersymmetry could allow
important achievements as well. Firstly, supersymmetry is an
 attractive concept  with
appealing possibilities in particle physics. The introduction of
local supersymmetry and subsquently of  supergravity  provide
an elegant gauge theory between bosons and fermions to which many hope
nature has reserved a rightful  place [2]. In fact,
N=1 supergravity is  a (Dirac) square root of gravity [3]:
physical states in the quantum theory must satisfy the
 supersymmetry constraints which
then imply with the quantum algebra that  the Hamiltonian constraints also
 to be satisfied  [3,4,5]. Secondly,
ultraviolet divergences  could be removed by
the presence of the extra symmetry [6].
Thirdly, it was suggested [7] that Planckian
effective masses  induced by wormholes
 could be eliminated with supersymmetry.

 Quite recently, an important result was achieved [8].
Namely, addressing the question of why the existence
of a Hartle-Hawking [9]  solution for Bianchi class A models
in pure N=1 supergravity [10-14]  seemed to depend on the
homogeneity condition for the gravitino [12].
In fact, it does not and it is now possible to find a Hartle-Hawking
and wormhole [15]  solutions in the same spectrum [8,43].
This result  requires    the inclusion of all allowed gravitational degrees of
freedom
into the Lorentz invariant fermionic sectors of the wave function.
However, there are many other issues in supersymmetric
quantum gravity which remain unsolved.
On the one hand,
why no physical states are found when a cosmological constant
is added [16-18] (nevertheless, a   Hartle-Hawking
solution was obtained for a $k=1$ FRW model) [Extending  the
framework presented in ref. [8] and using  Ashtekar variables,
it was shown in ref. [44] that the exponential of the Chern-Simons
functional constitute one case of solutions]   and
on the other hand, why the minisuperspace solutions
have no counterpart in the full theory because states
with zero (bosonic) or a finite number of fermions are not possible there [19].
A possible answer to the latter could be provided within the framework
presented in ref. [8]. But another problem has also been kept
without an adequate explanation: the apparent absence of
wormhole states either in some FRW [20,21] or
Bianchi IX models [22] when supermatter is included\footnote{$^{1}$}{\sevenrm
Other interesting issues in supersymmetric  quantum gravity/cosmology are:
a) obtaining conserved currents in minisuperspace from the wave function of the
universe,
$\Psi$ [23];
b) obtaining physical states in the full theory (are there any? how do they
look?) and
possibly checking
the conjecture made in [8]; c) why there are no physical states in a locally
supersymmetric
FRW model with gauged supermatter [24] but one can find them in  a locally
supersymmetric
FRW model with   Yang-Mills fields [25]. }. In addition, a
 Hartle-Hawking type solution can be found, even though some terms (which are
scalar field
dependent)
cannot be determined in a satisfactory way.

Classically, wormholes  join
different asymptotic regions of a Riemannian geometry.
Such solutions can  only be found when
certain types of matter fields are present
[15].
However, it  seems more natural to study quantum
wormhole states, i.e., solutions of the Wheeler-DeWitt
equation [15,26-29]. It is thought that wormholes may produce
 shifts  in  effective masses and interaction parameters [30,31].
Moreover, wormholes may play an important role
 which could force the cosmological constant
to be zero [32].
The wormhole ground state may be defined by a path integral over all possible
asymptotic
Euclidian 4-geometries and matter fields whose
energy-momentum tensor vanishes at  infinity.
Excited wormhole states would have sources at infinity.
However, the question concerning the main differences between a wormhole ground
state
and the excited states does not bear a simple answer.
In fact, if one has found the ground state (like in [15,36]) then excited
states
may be obtained from the repeated aplication of operators
(like ${\partial \over {\partial \phi}}$, e.g.) and
implementing their orthonormality. But it is another issue if one happens to
find a set of solutions from the Wheeler-DeWitt equation and tries to
identify which correspond to a   wormhole ground state or to  excited states.
Recent
investigations on this problem [26,28] claim that what may be  really
relevant is to use the whole  basis of wormhole solutions
(namely, to calculate the effects of wormhole physics from Green's functions,
where these have been factorized by introducing a {\it complete} set of
wormhole states [15])
and not just trying to identify and label a  explicit
expression which would correspond either to a wormhole  ground
state or an excited one.

The Hartle-Hawking (or no-boundary proposal) [1,9] solution is expressed in
terms
of a Euclidian path integral. It is essentially a topological statement about
the class of histories summed over. To calculate the no-boundary
wave function we are required to regard a three-surface as the {\it only}
boundary of a compact four-manifold, on which the  four-metric is
$g_{\mu\nu}$ and induces $h_{ij}^0$ on the boundary, and the matter field
is $\phi$ and matches $\phi_0$ on the boundary as well. We are then instructed
to
perform a path integral over all such $g_{\mu\nu}$ and $\phi$ within all such
manifolds.
For manifolds of the form of ${\bf R} \times \Sigma$, the no-boundary proposal
indicates us to choose initial conditions at the initial point as to ensure the
closure of the
four geometry. It basically consists in setting the initial three-surface
volume $h^{1/2}$ to
zero but also involve regular conditions on the derivatives of the remaining
components of the
three-metric and the matter fields [1,9].

Let me  briefly exemplify
how wormhole states seem to be absent and
why a Hartle-Hawking solution is only partially       determined.
Considering the more general theory
of N=1 supergravity with supermatter [33],
 one  takes
a k = + 1 FRW model with complex scalar fields
$\phi, \ol\phi$,
their  fermionic partners, $\chi_A$, $\ol\chi_{A'}$,
and   a  two-dimensional  spherically symmetric K\"ahler geometry.
The main results were shown {\bf not} to depend on the
fermionic derivative factor ordering and possible K\"ahler geometry
 [21].
Using the homogeneous FRW Ansatz for the fields (which for the
gravitino is $\psi^A_i=e^{AA'}_i\bar\psi_{A'}$ [35,36]), redefining
$  \rchi_{A} \rta a^{3 \over 2}  (1 + \phi \bar \phi)^{-1} \rchi_{A}$, $
\psi_{A} \rta  a^{3 \over 2} \psi_{A}$  to get simple Dirac brackets and using
instead
$ \bar \psi_{A} = 2 n_{A}^{~B'} \bar \psi_{B'}~, ~
 \bar \rchi_{A} = 2 n_{A}^{~B'} \bar \rchi_{B'} $
the supersymmetry
constraints are
$$ S_{A} = {1 \over \sqrt{2}} (1 + \phi \bar \phi) \rchi_{A} \pi_{\phi} - {i
\over 2 \sqrt{6}} a \pi_{a} \psi_{A}
 - \sqrt{3 \over 2} \sigma^{2}a^{2} \psi_{A} - {5i \over 4 \sqrt{2}} \bar \phi
\rchi_{A} \bar \rchi_{B} \rchi^{B} $$
$$+{1 \over 8 \sqrt{6}} \psi_{B} \bar \psi_{A}  \psi^{B} - {i \over 4 \sqrt{2}}
\bar \phi \rchi_{A} \psi^{B} \bar \psi_{B}  + {5 \over 4 \sqrt{6}} \rchi_{A}
\psi^{B} \bar \rchi_{B} + {\sqrt{3} \over 4 \sqrt{2}} \rchi^{B} \bar \rchi_{A}
\psi_{B}
 - {1 \over 2 \sqrt{6}} \psi_{A} \rchi^{B} \bar \rchi_{B}  \eqno(1)$$
and
its hermitian conjugate. Note that these expressions were
obtained directly from a canonical action of the form $\int dt (p\dot{q} -
H)$,
where $H={\cal N H} + \psi^A_0 S_A + \psi^{A'}_0 \ol S_{A'}$. $\cal N$ is the
lapse function.
Here, one  uses  $\hbar=1$ and $\sigma^2 = 2\pi^2$.
We choose $ (\rchi_{A} , \psi_{A} , a , \phi , \bar \phi) $ to be the
coordinates and
($ \bar \rchi_{A},$ $ \bar \psi_{A}, $ $ \pi_{a},$ $ \pi_{\phi}$ ,$ \pi_{\bar
\phi} $)
to be the momentum operators.

Some criteria have been presented to determine a suitable factor ordering.
This  problem  is related to the presence of cubic terms
in the supersymmetry constraints.
Basically,
$S_A, \ol S_A, {\cal H}$ could be chosen by requiring that [35,37]:

{\bf 1.}~ $S_A \Psi =0 $ describes the transformation properties of $\Psi$
under right handed supersymmetry
transformations (in the  ($a, \psi_A$) representation),

{\bf 2.}~ $\ol S_A \Psi =0 $ describes the transformation properties of $\Psi$
under left handed supersymmetry
transformations (in the  ($a, \ol \psi_A$) representation),

{\bf 3.}~ $S_A, \ol S_A$ are Hermitian adjoints with respect to an adequate
inner product [5],

{\bf 4.}~ A Hermitian Hamiltonian ${\cal H}$ is defined by consistency of the
quantum algebra.

However,  not all of these criteria can be
satisfied simultaneously (cf. [35,37]). An arbitrary choice is to satisfy {\bf
1,2,4} as in
here and [20,21,35,37,38]. Another possibility (as in [20,21,36]) is to
go beyond this factor ordering and insist that $S_A, \ol S_A$  could still
be related by a Hermitian adjoint operation (requirement {\bf 3.}). If one
adopts this
then there are some quantum corrections to $S_A, \ol S_A$ (namely, adding
terms linear in
$\psi_A, \chi_A$ to $S_A$ and linear in
$\ol \psi_A, \ol\chi_A$ to $\ol S_A$) which nevertheless modify the
transformation rules for the
wave function under supersymmetry requirements {\bf 1,2.}

Following the ordering used in ref.[20,21,35,37,38],
one  puts all the fermionic derivatives in  $S_{A}$ on the right. In $\ol S_A$
all the
fermionic derivatives are on the left.

The Lorentz constraint  $ J_{AB} = \psi_{(A} \bar
\psi_{B)} - \rchi_{(A} \bar \rchi_{B)} $
imply for  $\Psi$
$$ \Psi = A + iB \psi^{C} \psi_{C} + C \psi^{C} \rchi_{C} + iD \rchi^{C}
\rchi_{C} +
E \psi^{C} \psi_{C} \rchi^{D} \rchi_{D}~, \eqno (2)$$
where $A$, $B$, $C$, $D$, and $E$ are functions of $a$, $\phi$ and $\bar \phi$
only.
Using eq. (10) and its hermitian conjugate,   one   gets four equations from  $
S_{A} \Psi = 0 $
and  another four equations from $ \bar S_{A} \Psi = 0 $ (all first order
differential equations!)
which give
$$ A = f(\bar \phi) \exp({-3 \sigma^{2} a^{2}})~,~
 E = g(\phi) \exp ({3 \sigma^{2} a^{2}}) \eqno(3) $$
where $ f , g $ are arbitrary anti-holomorphic and
holomorphic functions of $\phi$, respectively.
Decoupling  the equations for $B,C,D$ (cf. ref. [21] for more details)
one finds
$$ B = h(\bar \phi) (1 + \phi \bar \phi)^{- {1 \over 2}} a^{3}
\exp({3 \sigma^{2} a^{2}})~,~
 C = 0 ~,~ D = k(\phi) (1 + \phi \bar \phi)^{- {1 \over 2}} a^{3} \exp ({-3
\sigma^{2} a^{2}} ) ~.
\eqno(4)$$
The result (4) is direct consequence that one could not find a consistent
(Wheeler-DeWitt type) second-order differential equation for $C$ and hence to
$B,D$.
It came directly from the corresponding first order differential equations.
Changing $S_A, \ol S_A$  in order that they can be related by some Hermitian
adjoint transformation
({\bf 3.})
gives essentialy
the same outcome [21]. With a two-dimensional flat
K\"ahler geometry one gets a similar result.

 While Lorentz invariance allows the pair $\psi_A\chi^A$ in (2),
supersymmetry  rejects it.
A possible interpretation could be that supersymmetry transformations
 forbid any fermionic bound state
 $\psi_A\chi^A$ by treating the  spin-$\half$ fields
 $\psi^A, \chi^B$ differently.

A Hartle-Hawking wave function\footnote{$^{2}$}{{\sevenrm
The Hartle-Hawking solution  could  not be  found
in the Bianchi-IX model of ref. [22].
Either a different homogeneity condition (as in [12]) for
$\psi^A_i$ or the framework of [8] could assist us in this particular
problem.}}
 could be identified in the fermionic filled sector, say,
$  g(\phi) \exp ({3 \sigma^{2} a^{2}})$, but for particular expressions of
$ g(\phi)$.  We notice though that the Lorentz and superymmetry constraints
 are not enough to specify
$ g(\phi)$. A similar situation is also present in
ref. [36], although an extra multiplicative factor of  $a^5$ multiplying
$ g(\phi)$ induces a less clear situation. In fact, no attempt was made in ref.
[36,38]
to obtain a Hartle-Hawking wave function solution. Being $N=1$ supergravity
considered
as a square root of general relativity [3], we would expect to be able
to find solutions of the type $e^{ik\phi} e^{a^{2}}$. These would correspond
to a FRW model with a massless minimally coupled scalar field in
ordinary quantum cosmology [1,41].

In principle, there are no physical arguments for wormhole
states to be absent in N=1 supergravity with supermatter.
In ordinary FRW quantum cosmology with scalar matter fields,
the wormhole ground state solution
would have a form like $e^{-a^2\cosh(\rho)}$,
where $\rho$ stands for a matter fields function
[15,26-28]. However,
such behaviour is not
provided by eqs.  (3), (4).
Actually, it seems quite different. Moreover, we may
ask in which conditions can these solutions  be accomodated in order for
wormhole type solutions to be obtained.
The arbitrary functions $f(\phi, \ol\phi), g(\phi, \ol\phi),
h(\phi, \ol\phi), k(\phi, \ol\phi)$  do not allow to conclude unequivovally
that in  these  fermionic
sectors the corresponding bosonic amplitudes
would be damped at   large 3-geometries for any allowed value of
$\phi, \ol\phi$ at  infinity. Claims were then made in ref. [20,21]
that no wormhole states could be found. The reasons were that the Lorentz and
supersymmetry constraints do  not seem  sufficient in this case to specify
the $\phi\ol\phi$ dependence of  $f, g, h, k$.

Hence, one  has a canonical formulation of N=1 supergravity which constitutes
a (Dirac) like square root of gravity [3,4,5].
Quantum wormhole and
Hartle-Hawking solutions were found in minisuperspaces for
pure N=1 supergravity [8,10-14,17,18,34-35,37] but
the former state is absent in the literature
\footnote{$^{3}$}{{\sevenrm
Notice that for pure gravity neither classical or quantum wormhole
solutions have been produced in the literature. A matter field seems to be
required: the ``throat'' size is proportional
to $\sqrt{\kappa}$ where $\kappa$ represents   the (conserved) flux of
matter fields.}},
for pure gravity cases
[1,9,15,26-28]. Hartle-Hawking  wave functions and wormhole ground states
are present in
ordinary minisuperspace with matter
[1,9,15,26-28]. When supersymmetry is introduced [20-22,35-38]
one faces some problems
within the more general theory of N=1 supergravity with
supermatter [33]  (cf. ref. [20-22]) as far as Hartle-Hawking or wormhole type
solutions are concerned.
An attempt [38] using the constraints present in [35,37] but the
ordering employed above, also  seemed to have failed  in getting wormhole
states. In addition, a model
combining a conformal scalar field with spin-$\half$ fields
(expanded in  spin$-\half$ hyperspherical harmonics and
integrating over the spatial coordinates [30]) did not produce any
wormhole solution as well [39].
However, ref. [36] clearly represents an opposite point of view, as it
explicitly depicts wormhole ground states in a locally supersymmetric setting.

It might be interestig  to point  that the
constraints employed in [36]
(and also in [35,37,38]) were derived from a {\it particular}
model constructed in [40], while ours [21] come directly from
the {\it more general} theory of N=1 supergravity coupled to
supermatter [33].
 Moreover, there are many differences
between the expressions in [34=37] and the one  hereby (see also [21]),
namely on numerical coefficients.

Let me sketch briefly how the supersymmetry constraints expressions in
[36] were obtained. First, at the pure N=1 supergravity level,
the following re-definition of fermionic non-dynamical variables
$$ \rho^A \sim a^{-1/2}\psi^A_0 +
{\cal N} a^{-2} n^{AA'}\overline{\psi}_{A'}, \eqno(5)$$ and its hermitian
conjugate
were introduced for a FRW model, changing the supersymmetry and Hamiltonian
constraints.
As a consequence, no
fermionic terms were present in  ${\cal H} \sim \{S_A, \ol S_A\}$ and no
cubic fermionic terms in the supersymmetry constraints. Hence,
no ordering problems with regard to fermionic derivatives were present. The
model with matter
was then extracted {\it post-hoc} [35,37] from a few basic assumptions
about their general form and supersymmtric algebra. This simplified route
seemed
to give similar expressions, up to minor field redefinitions,
to what one would obtain for a reduced model from the {\it particular}
theory presented in [40], as stated in
[35,37]. Note that cubic fermionic terms
like $\psi\ol\psi\psi$ or $\psi\ol\chi\chi$ are now present but the former is
absent
in the pure case.
In ref. [35,37,38], criteria {\bf 1,2,4} were used for the fermionic derivative
ordering, while
in ref. [36] one insisted to accomodate an Hermitian adjoint relation
between the supersymmetry trnasformations ({\bf 3.}). It so happens that a
wormhole
ground state was found in the former but not in the latter. In ref. [20,21]
the same possibilites for using these criteria  were employed but
with supersymmetry and Hamiltonian constraints directly obtained from
$\psi^A_0, \ol \psi^{A'}_0, {\cal N}$ (see eq. (5)). Apparently, no wormhole
states were present. Moreover, we also recover a solution which satisfied only
partially
the no-boundary proposal conditions (see eq. (3)). A similar but yet less clear
situation also seems to be present in ref. [36].

The issue concerning the existence or not of  wormhole and
Hartle-Hawking quantum cosmological states for
 minisuperspaces within N=1 supergravity with supermatte is therefore of
relevance [42].
The current literature on the subject is far from a consensus.
No explanation has been provided for the
(apparent) opposite conclusions [20,21,33] concerning the
existence of wormhole   states and to point out which is right and why.
Furthermore, it does not seem possible for
the procedure presented  in [8] to solve
this conundrum.

Here  an answer for this particular problem is presented.
The explanation   is that chosing the type of Lagrange multipliers
and the fermionic derivative ordering one uses
 makes a
difference.  Our arguments are as follows.

On the one hand, the quantum formulation of wormholes in ordinary
quantum cosmology has been shown to depend on the lapse function [27,28].
Such ambiguity has already been pointed out
in [41] (see also [45]) but for generic quantum cosmology and related to
bosonic factor ordering questions in the Wheeler-DeWitt operator.
An ordering is necessary in order to make predictions. A proposal was made
that the kinetic terms in the Wheeler-DeWitt operator should be the Laplacian
in the natural (mini)superspace element of line, i.e., such that it would
be invariant under changes of coordinates in minisuperspace [41]. Basically,
this includes the  Wheeler-DeWitt operator to be locally self-adjoint in the
natural
measure generated by  the above mentioned element of line.
However, it suffers from the problem that the connection defined by a
minisuperspace
line element like $ds^2 = {1 \over {\cal N}} f_{\mu\nu} dq^\mu dq^\nu$ could
not be linear
on $\cal N$. This would then  lead to a Wheeler-DeWitt operator {\it not}
linear in $\cal N$
as it would be  in order that $\cal N$ be interpreted as a Lagrange multiplier
(it was also proposed in ref. [41] that this possible non-linearity dependence
on $\cal N$ could cancel out in theories like supergravity where bosons and
fermions
would be in equal number of degrees of freedom).
For each choice of $\cal N$, there is a different metric in minisuperspace, all
these
metrics being related by a conformal transformation [46]. Therefore, for each
of these
choices, the quantization process will be different. In fact, for a
minisuperspace consisting
of a FRW geometry and homogeneous scalar field, a conformal coupling allows a
more
general class of solutions of the Wheeler-DeWitt equation than does the
minimally
coupled case, even if a one-to-one correspondence exists between bounde states
[46].

For some choices of $\cal N$ the quantization are
 even inadmissible,
e.g,
when ${\cal N} \rta 0$ too fast for vanishing
3-geometries in the wormhole case.
Basically, requiring regularity for $\Psi$ at $a\rta 0$ is equivalent
to self-adjointness for the Wheeler-DeWitt operator at that point. Such
extension
would be expected since wormhole wave functions calculated via a path integral
are
regular there. Three-geometries with zero-volume would be a consequence of the
slicing procedure which has been carried. In other words, $a=0$ simply
represents a coordinate singularity in minisuperspace. An extension
for (and beyond it), similar to the case of the Rindler wedge and the full
Minkowski space,
would be desirable. The requirement that the Wheeler-DeWitt operator be
self-adjoint
selects a scalar product and a measure in minisuperspace. Gauge choices of
$\cal N$
that vanish too fast when $a \rta 0$ will lead to problems as the
minisuperspace measure will be infinite at (regular) configurations associated
with vanishing three-geometries volume.
The difference on the quantization
manifests itself in the Hilbert space structure of the wormhole
solutions due to the scalar product dependence on $\cal N$ and not in
the structure of
the Wheeler-DeWitt operator or path integral.
More precisely, the formulation of
global laws, i.e., finding boundary conditions
for the Wheeler-DeWitt equation in the wormhole case, equivalent
to the ones in the path integral approach,
could depend
on the choice of $\cal N$ but not the
local laws in minisuperspace\footnote{$^{3}$}{{\sevenrm
Physical results such as effective interactions are independent of the
choice of $\cal N$ due to the way the corresponding path integrals  are
formulated.}}.

On the other hand, a similar effect seems to occur when local supersymmetry
transformations
are present. Besides the lapse function, we have now the
time components of the gravitino field, $\psi^A_0$, and
of the torsion-free connection $\omega_{AB}^0$ as Lagrange multipliers.
If one uses transformation (5) but without the last term, then
the supersymmetry and Hamiltonian constraints read  (in the pure case):

$$ S_A = \psi_A\pi_a - 6ia \psi_A +
{{i}\over{2a}}n_A^{E'}\psi^E\psi_E\overline{\psi}_{E'}, \eqno(6a)$$

$$ \overline{S}_{A'} = \overline{\psi}_{A'} \pi_a +
6ia \overline{\psi}_{A'} - {{i}\over{2a}}n_E^{A'}\overline{\psi}^{E'}
\psi_E\overline{\psi}_{E'},  \eqno(6b)$$

$$ {\cal H} = -a^{-1} (\pi_a^2 + 36 a^2)       + 12a^{-1} n^{AA'}\psi_A
\overline{\psi}_{A'}. \eqno(6c)$$

If $\rho_A, \ol \rho_{A'}$ had been used instead of $\psi^A_0, \ol \psi^{A'}_0$
then the
second terms in (6a)-(6c) would be absent. I.e.,
for the transformation  (5) the corresponding supersymmetry constraints and the
Hamiltonian are
either linear or free of fermionic terms (cf. eq. (1) and ref. [34,35,37] as
well). What seems to
have been gone  unnoticed is the following.
{\it Exact} solutions of $S_A \Psi =0$ and $\ol S_A \Psi =0$ (using the
criteria {\bf 1,2,4})
in the pure case for (6a),(6b) with or without second term
are  $A_1 = e^{-3a^2}$ and $A_2 = e^{3a^2}$, respectively, for
$\Psi = c A_1 + d A_2 \psi_A\psi^A$ where $c, d$ are constants.
This $\Psi$ represents a linear combination of
of WKB solutions of
${\cal H}\Psi =0$, obtained form the corresponding Hamilton-Jacobi equation,
i.e.,
they represent a {\it semi-classical approximation},
{\it but only} for the $\cal H$ without the second term in (6c), i.e.,
when (5) is fully employed. Strangely it does not for the full expression in
(6c); in fact the function $e^{3a^2}$ would have to be replaced.

Hence the choice between $\rho_A$ and $\psi^A_0$ directly affects any
consistency between the
quantum solutions of the constraints (6a)-(6c). Moreover, an important point
(which will be stressed later) is that the Dirac-like equations in ref. [36]
lead consistently to a set of Wheeler-DeWitt equations (like in [35,37,38])
but that could not be entirely achieved in ref. [20,21].
As explained in eq. (4), the difficulty in determining the
$\phi, \ol\phi$ dependence of $f,g,h,k$ (and therefore to acess on the
existence of wormhole states) is related to the fact that $C=0$, which
is an indication as well that  corresponding Wheeler-DeWitt equations
could not be obtained from the supersymmetry constraints.

Choosing (5) one achieves the simplest form for
the supersymmetry and Hamiltonian constraints and their
Dirac brackets. This is important at the pure case level,
as far as the solutions of  $S_A \Psi =0$ and $\ol S_A \Psi =0$
are concerned. Moreover, fermionic factor ordering become absent in that case.
If one tries to preserve this property through a {\it post-hoc}
approach [35,37] when going to the matter case (keeping a simplified
form for the constraints and algebra) then one might hope to avoid any problems
like
the ones refered  to in eq. (4). In addition, using the fermionic ordering of
[36]
where one accomodates the Hermitain adjointness with {\bf 1,2,4} up to minor
changes relatively to {\bf 1,2}, one does get a wormhole groud state. Thus,
there seems to be a relation between a choice of Lagrange multipliers (which
simplifies the constraints and the algebra in the pure case), fermionic factor
ordering
(which may become absent in the pure case) and obtaining from the
supersymmetry constraints second order consistency equations (i.e.,
 Wheeler-DeWitt type equations). The failure of this last one  is the
reason why $C=0$ and $f,g,h,k$ cannot be determined from the algebra.
Different choices of $\psi^A_0$ or $\rho_A$, then of fermionic derivative
ordering will lead to
different supersymmetry constraints and to different solutions for the
quantization of the problem.
It should also be stressed that from the supersymmetric algebra a combination
of
two supersymmetry
transformations, generated by $S_A$ and $\ol S_{A'}$ and whose amount is
represented
by the Lagrange multipliers $\psi^A_0, \ol \psi^{A'}_0$, will be (essentially)
equivalent to a
transformation generated by
the Hamiltonian constraint and where the lapse function is the corresponding
Lagrange multiplier.

So, how should the search for wormholes ground
states\footnote{$^{5}$}{{\sevenrm
Regarding the Hartle-Hawking solutions  it seems it can be obtained
straightforwardly either
up to a specific definition of homogeneity [12] or following the approach in
[8]. This might help
in regarding the results found in [22] with respect to the Hartle-Hawking
solution.}}
in N=1 supergravity be approached?
One possibility would be to  employ  a
transformation like (25) (see [35]). In fact, using it from the begining in our
case model it will
change some coefficients in the supersymmetry constraints as it can be
confirmed.
As a consequence, we are then allowed to get consistent second order
differential equations
from $S_A \Psi =0$  and $\ol S_{A'} \Psi =0$. Hence, a line equivalent
to the one followed in ref. [36] can be used and a wormhole ground state be
found.
Alternatively, we could
restrict to the {\it post-hoc} approach introduced and
followed throughout in [34,38] as explained above. Another possibility, is to
extend the approach introduced by L. Garay [26-29] in ordinary quantum
cosmology
to the cases where local supersymmetry is present.
The basic idea is that what is really relevant is to determine a {\it whole}
basis
of wormhole solutions of  the associated Wheeler-DeWitt operators,
not just trying to identify  one single solution like the ground state
from a all set of solutions. Hence, one
ought to adequatly define what a basis of wormhole solutions means.
In this case, we could be able to still use any Lagrange multiplier (just as
$\psi^A_0$),
avoiding having to find a  redefinition of fermionic variables  as in (5) but
for the matter case in question (scalar, vector field, etc).

Basically, improved boundary conditions for wormholes can be formulated
by requiring square integrability in the {\it maximmaly extended}
minisuperspace
[27,28]. This condition ensures that $\Psi$ vanishes at the {\it truly}
singular configurations
and guarantees its regularity at any other (coordinate) one,
including vanishing 3-geometries.
A   maximally extended minisuperspace and a proper
definition of its boundaries in order to
comply with  the behaviour of $\Psi$ for $a\rta 0$ and
$a\rta \infty$ seems to be mandatory  in ordinary
quantum gravity. The reason was that the
quantum formulation of wormholes has been shown to depend on
the lapse function, $\cal N$ [26,28].
The maximal analytical extension of minisuperspaces can be considered as the
natural configuration space for quantization [26]. The boundary of the
minisuperspace
would then consist of all those configurations which are truly singular.
Any  regular configurations will be in  its interior.
Another reason to consider the above boundary conditions in a
maximally extended minisuperspace is that it allow us to avoid boundary
conditions at
$a=0$ to guaratee the self-adjointness of the Wheeler-DeWitt operator.
This operator is hyperbolic and well posed boundary conditions
can only be imposed on its characteristic surfaces and the one associated with
$a=0$ may not be of this type, like in the case of a conformally coupled scalar
field.
In such a case, it would be meaningless to require self-adjointness there
(cf. ref. [26,28] for more details).

Within  this framework wormhole solutions would form a Hilbert space.
These ideas  must then be extended    to
a case of locally supersymmetric minisuperspace with
odd Grassmann (fermionic) field variables.
In this case, not only one has to deal with different
possible behaviours for  $\cal N$ but also with $\psi^A_0$.
Then, it will be possible to determine
explicitly the form of $f,g,h,k$ in order that some or even an overlap of them
could provide
  a wormhole wave function, including the ground state. In fact, this would
mean that not only
the bosonic amplitudes $A,B,..$ would
have to be considered for solutions
but the fermionic pairs ought to be taken
as well.
Constructing an adequate Hilbert space from (3),(4) would
lead us to a basis of wormhole states in such a
singularity-free space (see [26]). Wormhole wave functions
could be interpretated in terms of overlaps between
different states.

Another point which might be of some relevance
is the following [28]. The  evaluation of the path integral
(or say, determining the boundary conditions for the Wheeler-DeWitt equation)
for
wormhole states in ordinary minisuperspace quantum cosmology requires the
writing
of an action adequate to asymptotic Euclidian space-time, through the inclusion
of
necessary boundary terms [15,26-28]. There may
changes when fermions and supersymmetry come
into play. A {\it different} action\footnote{$^{6}$}{{\sevenrm
The canonical form of action of
pure N=1 supergravity present in the literature [5]
(which includes boundary terms)
 is  not
invariant under supersymmetry transformations.
Only recently a fully invariant action but restricted to
 Bianchi class A models was presented [14].}} would then
induces improved  boundary conditions for the intervening
fields
as far a wormhole Hilbert space structure is concerned
in a locally supersymmetric minisuperspace.

Summarizing,  the issue
 concerning the existence
of wormhole states in locally supersymmetric minisuperspace
models was addressed in this work.
Wormhole  states
are apparently  absent in models obtained from
the more  general theory of N=1 supergravity with supermatter.
As explained, the cause
 investigated here is that an appropriate  choice of
Lagrange multipliers and fermionic derivative makes a difference.
 From the former we get the simplest form of the supersymmetry and
Hamiltonian constraints and their Dirac brackets in the pure case. This ensures
no
fermionic derivative ordering problems and that the solutions of the quantum
constraints are consistent.
Either from a {\it post-hoc} approach (trying to extend the  obtained framework
in the pure case)
or from a direct dimensional-reduction we  get consistent second order
Wheeler-DeWitt type
equations or corresponding solutions     in the supermatter case.
 From an adequate use of criteria {\bf 1,2,3,4} above, we get a wormhole ground
state.
We also notice  that the  use of appropriate Lagrange multipliera also requires
a specific
fermionic ordering results in order  to
 obtain a consistency set of  Wheeler-DeWitt equations or respective solutions.
The search for wormhole solutions could also be addressed from another
point of view [28,30]. One has to invest supersymmetric quantum wormholes
with a Hilbert space structure, associated with
a maximal analytical extension of the corresponding
minisuperspace.
A basis of wormhole states might  then be obtained from the many
possible solutions of the supersymmetry constraints  equations.

Finally, I would like to quote the following words from
C. Dickens book, ``A Tale of Two Cities'':

\vskip .05 true in

{\leftskip = 1.5in   {\it It was
the best of times, it was the worst of times; it was the age of wisdom, it was
the age of
foolishness; it was the epoch of belief, it was
the epoch of incredulity; it was the season of Light, it was the season of
Darkness;
it was the spring of hope, it was the winter of despair; we had everything
before us, we had
nothing before us...}
   \par}

Im my own opinion, it closely
describes most of the path followed by some of us and which still remains ahead
in
the subject of supersymmetric
quantum gravity/cosmology. Indeed, much more remains to be done in order
to properly accomodate
all basic   results and avoid any  paradoxical situations.

\noindent
{\bf Note added}

After completion of this work and before send it to the publishers, the author
received a paper [43] by A. Csord\`as and R. Graham. There, the problem
of a cosmological constant in supersymmetric minisuperspaces from N=1
supergravity
was dealt with and a solution proportional to exponential
of the Chern-Simons functional was found.

\noindent
{\bf ACKNOWLEDGEMENTS}

The author is  grateful to
A.D.Y. Cheng, L. J. Garay,  S.W. Hawking and O. Obregon for  helpful
conversations
and for sharing their points of view.
The author  would like to thank the Organizing Committee of the
VI Moskow Quantum Gravity International Semimar, Moskow, Russia, 12-19 June
1995,
for providing a delightful and stimulating atmosphere.
Questions and discussions with
 I. Antoniadis,
G. Esposito, B. Ovrut, D. Page, D. Salopek,  K. Stelle, A. Zhuk which
motivated
further improvements in the paper are also acknowledged.
This work was supported  by
a Human Capital and Mobility (HCM)
Fellowship from the European Union (Contract ERBCHBICT930781).

\noindent
{\bf REFERENCES}

{\rm

\advance\leftskip by 4em
\parindent = -4em

[0] G. Yeats, {\it Cambridge Colleges Ghosts}, Jarrold (Norwich, 1994).

[1] See for example, G. Gibbons and S.W. Hawking, {\it Euclidian Quantum
Gravity},
World Scientific (Singapore, 1993);

J. Halliwell, in: {\it Proceedings of the Jerusalem Winter School on
Quantum Cosmology and Baby Universes}, edited by T. Piram et al, World
Scientific, (Singapore, 1990)
 and refereces therein.
 and refereces therein.

[2]  P. van Nieuwenhuizen, Phys.~Rep. {\bf 68} (1981) 189.

[3] C. Teitelboim, Phys.~Rev.~Lett.~{\bf 38}, 1106 (1977), Phys. Lett. B{\bf
69} 240 (1977).

[4] M. Pilati, Nuc. Phys. B {\bf 132}, 138 (1978).

[5] P.D. D'Eath, Phys.~Rev.~D {\bf 29}, 2199 (1984).

[6] G. Esposito, {\it Quantum Gravity, Quantum Cosmology and Lorentzian
Geometries},
Sprin\-ger Verlag (Berlin, 1993).

[7] S.W. Hawking, Phys. Rev. D{\bf 37} 904 (1988).

[8]  R. Graham and A. Csord\'as, Phys. Rev. Lett. {\bf 74} (1995) 4129

[9] J.B. Hartle and S.W. Hawking, Phys.~Rev.~D {\bf 28}, 2960 (1983).

[10] P.D. D'Eath, S.W. Hawking and O. Obreg\'on, Phys.~Lett.~{\bf 300}B, 44
(1993).

[11] P.D. D'Eath, Phys.~Rev.~D {\bf 48}, 713 (1993).

[12] R. Graham and H. Luckock, Phys. Rev. D
{\bf 49}, R4981 (1994).

[13] M. Asano, M. Tanimoto and N. Yoshino, Phys.~Lett.~{\bf 314}B, 303 (1993).

[14] H. Luckock and C. Oliwa,
(gr-qc 9412028), accepted  in
Phys. Rev. D.

[15] S.W. Hawking and D.N. Page, Phys.~Rev.~D {\bf 42}, 2655 (1990).

[16] P.D. D'Eath, Phys. Lett. B{\bf 320}, 20 (1994).

[17]  A.D.Y. Cheng, P.D. D'Eath and
P.R.L.V. Moniz, Phys. Rev. D{\bf 49} (1994) 5246.

[18]  A.D.Y. Cheng, P. D'Eath and
P.R.L.V. Moniz,
{\rm Gravitation and Cosmology}
{\bf 1} (1995) 12

[19] S. Carroll, D. Freedman, M. Ortiz and
D. Page, Nuc. Phys. B{\bf 423}, 3405 (1994).

[20]  A.D.Y. Cheng, P.  D'Eath and
P.R.L.V. Moniz,
{\rm Gravitation and Cosmology}
{\bf 1} (1995) 1

[21]  A.D.Y. Cheng and P.R.L.V. Moniz,
Int. J. Mod. Phys. {\bf D4}, No.2 April (1995) - to appear.

[22]   P. Moniz,
{\it Back to basics? ... or How can supersymmetry be used in a simple
           quantum cosmological model}, communication
presented at 1st Mexican School on Gravitation and
           Mathematical Physics, Guanajuato, Mexico, 12-16 Dec 1994,
gr-qc/9505002;

{\it Quantization of the Bianchi type-IX
model in N=1  Supergravity in the
presence of supermatter}, DAMTP report R95/21, gr-qc/9505048, submitted to
International
Journal of Modern Physics {\bf A}.

[23] J. Bene and R. Graham, Phys. Rev. {\bf D49} (1994) 799; R. Mallett, Class.
Quantum Grav. {\bf 12} (1994) L1; A. Cheng, O. Obregon and P. Moniz, in
preparation.

[24]  A.D.Y. Cheng, P.D. D'Eath and
P.R.L.V. Moniz, DAMTP R94/44, Class. Quantum Grav. to appear.

[25] P. Moniz, {\it Physical States in a Locally Supersymmetric FRW model
coupled
to Yang-Mills fields}, DAMTP Report, in preparation

[26] L. Garay, Phys. Rev. D{\bf 48 } (1992) 1710.

[27] L. Garay, Phys. Rev. D{\bf 44} (1991) 1059.

[28] L. Garay, Ph. D. Thesis (in spanish) Madrid - Consejo Superior de
Investigaciones
Cientificas, 1992.

[29] G. Mena-Marugan, Class. Quant. Grav. {\bf 11} (1994) 2205; Phys. Rev.
D{\bf 50} (1994) 3923.

[30] A.Lyons, Nuc. Phys. {\bf B}324 (1989) 253.

[31]  H.F. Dowker , Nuc. Phys. {\bf B}331 (1990) 194;
 H.F. Dowker and R. Laflamme, Nuc. Phys. {\bf B}366 (1991) 201.

[32] S. Coleman, Nuc. Phys. {\bf B}310 (1988) 643.

[33]  J. Wess and J. Bagger, {\it Supersymmetry and Supergravity},
2nd.~ed. (Princeton University Press, 1992).

[34] P.D. D'Eath and D.I. Hughes, Phys.~Lett.~{\bf 214}B, 498 (1988).

[35] P.D. D'Eath and D.I. Hughes, Nucl.~Phys.~B {\bf 378}, 381 (1992).

[36] L.J. Alty, P.D. D'Eath and H.F. Dowker, Phys.~Rev.~D {\bf 46}, 4402
(1992).

[37] D.I. Hughes, Ph.D.~thesis, University of Cambridge (1990), unpublished.

[38] P. D Eath, H.F. Dowker and D.I. Hughes, {\it Supersymmetric
Quantum Wormholes States} in: Proceedings of the Fifth Moskow Quantum Gravity
Meeting, ed. M. Markov, V. Berezin and V. Frolov, World Scientific (Singapore,
1990).

[39] H.F. Dowker, Ph.D. Thesis, chapter 4, University of Cambridge (1991),
unpublished

[40] A. Das. M. Fishler and M. Rocek,  Phys.~Lett.~B {\bf 69}, 186 (1977).

[41] S.W. Hawking and D. Page, Nuc. Phys. B{\bf 264} (1986) 185.

[42] P. Moniz, {\it Is there a problem with quantum wormholes in N=1
supergravity?},
DAMTP R95/19, to be submitted to General Relativity and Gravitation

[43] A. Csord\`as and R.   Graham, {\it  Nontrivial fermion states in
supersymmetric minisuperspace},
talk presented at Mexican School in Gravitation and
           Mathematical Physics, Guanajuato, Mexico, Dec 12-16, 1994,
gr-qc/9503054

[44] A. Csord\`as and R.   Graham,, {\it Quantum states on supersymmetric
minisuperspaces with
cosmological constant}, gr-qc/9506002

[45] T. Padmanabhan, Phys. Rev. Lett. {\bf 64} (1990) 2471

[46] D. Page, J. Math. Phys. {\bf 32} (1991) 3427

\ \ \ }

\bye
\end